\documentclass[twocolumn, tighten, times, twocolappendix, longbib,
]{aastex701}
\usepackage{amsmath}

\shorttitle{Chemical enrichment of metal-poor stars orbiting massive black hole companions \ \ }
\shortauthors{Rosselli-Calderon et al.}

\begin{document}
\title{Chemical enrichment of metal-poor stars orbiting massive black hole companions}

\author{Alejandra~Rosselli-Calderon}
\affiliation{Department of Astronomy and Astrophysics, University of California, Santa Cruz, CA 95064, USA}
\email{aleroca@ucsc.edu}

\author{Julia Stewart}
\affiliation{Department of Astronomy and Astrophysics, University of California, Santa Cruz, CA 95064, USA}
\affiliation{Astrophysics Program, the Graduate Center, City University of New York, 365 5th Ave, New York, NY 10016, USA}
\email{jurstewa@ucsc.edu}

\author{Sijing Shen}
\affiliation{Institute of Theoretical Astrophysics, University of Oslo, PO Box 1029 Blindern, 0315 Oslo, Norway}
\email{sijing.shen@astro.uio.no}

\author{Sukanya Chakrabarti}
\affiliation{Department of Physics and Astronomy, University of Alabama in Huntsville, 301 North Sparkman Drive, Huntsville, AL 35816, USA}
\email{sc0236@uah.edu}

\author{Melinda Soares-Furtado}
\affiliation{Department of Astronomy, University of Wisconsin-Madison, 475 N. Charter St., Madison, WI 53706, USA}
\email{mmsoares@wisc.edu}

\author{Enrico~Ramirez-Ruiz}
\affiliation{Department of Astronomy and Astrophysics, University of California, Santa Cruz, CA 95064, USA}
\email{enrico@ucolick.org}

\begin{abstract}
There are millions of undetected black holes wandering through our galaxy. Observatories like {\it Chandra}, LIGO, and more recently, {\it Gaia}, have provided valuable insights into the configurations of these elusive objects when residing in binary systems. Motivated by these advances, we study, for the first time,  the enhanced accretion of metals from the interstellar medium (ISM) onto low-mass companions in binary systems with highly unequal mass ratios, utilizing a series of hydrodynamical simulations. Our study demonstrates that a stellar companion's metal accretion history from the ISM alone, from its formation to the present, can significantly influence its surface abundance, especially when enhanced by a massive black hole companion. However, this effect is likely only measurable in stars that are still in the main sequence. Once a stellar companion evolves off the main sequence, similar to what has been observed with the {\it Gaia} BH3 companion, the initial dredge-up process are likely to erase any excess surface abundance resulting from the metals that were accreted. As we discover more unequal mass ratio binary systems, it is crucial to understand how the observed metallicity of sun-like companions may differ from their birth metallicity, especially if they are not yet evolved.
\end{abstract}


\section{Introduction}

In a universe where multiple-star systems are ubiquitous, it becomes essential to understand how the interactions between the companions shape their properties. For the case of massive stars, it is estimated that most of them reside in binaries \citep[e.g.,][]{Sana2012}. Recent observational studies additionally estimate that $70$\% of stars with masses of $M \approx 10\,M_{\odot}$ are found in multi-body systems \citep{Moe2019}. Moreover, there is an inverse correlation  between stellar metallicity and binary fraction,  implying that a larger fraction of lower metallicity stars are found in multiple stellar systems. It has been shown that solar-like stars with metallicities [Fe/H] lower than $-0.5$ have a multiplicity fraction $\approx 2-3$ times larger than that of solar metallicity stars \citep{2018ApJ...854..147B}. 

Stars are thought to retain a chemical record of their original environments through their current metal abundances \citep[e.g.,][]{Frebel_2013, Frebel_2015}. This characteristic makes low-mass, metal-poor stars (i.e., old stars) in the Galaxy important for studying star formation during the early epochs of the Universe \citep[e.g.,][]{2018ApJ...860...89M}. These stars provide valuable insights into the conditions that existed during the Galaxy's formation and may reveal characteristics of the earliest metal-poor stars \citep{Johnson_2015}, which are believed to be the stellar progenitors of black holes observed by  LIGO \citep[e.g.,][]{Kinugawa_2014}. 

Efforts to analyze the abundance properties of metal-poor stars depend on the assumption that these stars accurately reflect the metal abundance of the star-forming clouds in which they were born \citep[e.g.,][]{2022ApJ...936L..26K,2023ApJ...949..100K}. However, this assumption may be incorrect if stars are able to efficiently accrete metals from the interstellar medium (ISM) as they orbit within the Galaxy's gravitational potential over their lifetimes \citep[e.g.,][]{Bondi1944,Talbot_1977}. Given their significant ages, low-mass, metal-poor stars could interact with a substantial volume of ISM gas, depending on their orbital evolution. Over time, the ISM itself becomes enriched with metals produced by other stars \citep{2018MNRAS.477.1206N}.

The chemical enrichment of stars through accretion from the ISM has been the subject of study for several decades \citep{Talbot_1977, Alcock_1980, Yoshii_1981, Iben_1983, Frebel_2009, Komiya_2009, Komiya_2010, Johnson_2011, Hattori_2014, Johnson_2015, Komiya_2015}.  Notably, using the Eris zoom-in cosmological simulation \citep{2011ApJ...742...76G,2015ApJ...807..115S}, which models the assembly of a Milky Way analog, \citet{Shen_2017} explore the chemical enrichment of sun-like stars (with shallow convective layers) due to the accretion of metal-enriched gas from the ISM during the Galaxy's development. They concluded that a metal-poor star traveling through the ISM could accumulate enough material throughout its lifetime to potentially increase its birth metallicity. In extreme cases, accretion onto these relatively pristine stars can reach enrichment levels of [Fe/H]\,$\lesssim-2$, with the median enrichment level being [Fe/H]\,$\approx-5$   \citep{Shen_2017}.

Stellar multiplicity offers a unique opportunity to further investigate how the accretion of metal-enriched gas from the ISM can be enhanced by the combined gravitational pull in multiple star systems, particularly those that host massive black hole companions. Most known black holes are identified by the X-rays they emit while consuming material from their stellar companions \cite[e.g.,][]{2006ARA&A..44...49R}, known as high-mass
X-ray binaries (HMXBs). When black holes are more widely separated from their companions, they produce little to no radiation, making them detectable primarily through the gravitational influence they exert on their companions \citep{2019ApJ...886...68A,2019Sci...366..637T}. Remarkably, \textit{Gaia} has enabled the detection of three quiescent black holes \citep{2023MNRAS.518.1057E,2023MNRAS.521.4323E,2023AJ....166....6C,gaia_bh3}. 

In our Milky Way, most stellar-origin black holes have a mass of about ten solar masses \citep[e.g.,][]{2010ApJ...725.1918O}, with the previous record holder being the Cyg X-1 black hole, estimated to have a mass of around twenty solar masses \citep{2021Sci...371.1046M}. {The heaviest black hole detected in our Milky Way is {\it Gaia} BH3  \citep{gaia_bh3}, which has a mass of $32.7 \pm 0.82\,M_\odot$, placing it within the mass range observed by LIGO \citep{2023PhRvX..13a1048A}. The companion to this black hole is a $0.8 M_{\odot}$ G type star with a metallicity of $-2.56 \pm 0.11$. Remarkably, the stellar companion has a metallicity lower than that of $99.99\%$ of stars in the solar neighborhood, with only $1/10^4$ stars having a lower value \citep{2024OJAp....7E..38E}.} The metal-poor companion star in the \textit{Gaia} BH3 system supports the idea that massive black holes form efficiently in metal-poor environments early in cosmic history. While based on a single system, this observation raises the intriguing possibility that black hole companions may be preferentially found among metal-poor stars—a trend that future discoveries will be able to test. {Furthermore, stars in this mass regime could belong to the ``surviving stars" population, a class low mass population III formed in early hydrogen deuteride-cooling clouds \citep{nishijima2023lowmasspopiiistar}. }

Previous analytical and numerical studies of binary systems have primarily focused on cases with equal mass ratios \citep{Antoni2019,2019MNRAS.490.5196C,2023ApJ...944...44K}. In this work, we explore, for the first time, the evolution of binaries with unequal mass ratios to understand how their combined gravitational influence on the surrounding gas can significantly increase its original metallicity. This framework is essential for studying the ecosystems of binary systems and the complexities of their interactions. 

In this study, we extend both the analytical and numerical frameworks of classical Bondi-Hoyle-Lyttleton (BHL) accretion to binary systems in gaseous media \citep{Bondi1944, hoyle_lyttleton_1939}, specifically focusing on mass ratios where $M_1/M_2 \ll 1$. We conduct three-dimensional hydrodynamic simulations with a uniform-density background wind to investigate how the presence of a more massive binary companion affects the accretion rates experienced by the lighter member. Additionally, we integrate our accretion model with the stellar trajectories from the Eris zoom-in cosmological simulation \citep{Shen_2017} to analyze the chemical enrichment of low-mass stars  in binaries that have massive black hole companions, such as those discovered by \textit{Gaia}. This paper is structured as follows. In Section \ref{sec:motivation} we present the analytical formalism to frame our thinking and in Section \ref{sec: hydro} we introduce our  hydrodynamical framework and computational setup. In Section \ref{sec: results} we present the results of our numerical experiments and in Section \ref{sec: case study} we use the case study of \textit{Gaia} BH3 as an illustration of the effects of accretion on low-mass stars when having  massive black hole companions. Section \ref{sec: conclusions} summarizes our findings and the implications that the idealized results have on high-mass-ratio binaries. 

\section{Accretion enhanced by A massive binary companion}\label{sec:motivation} 
Binary stars often travel through the ambient medium from which they can accrete material. In such a context, the rate at which matter is accreted depends on the properties of the binary system, such as the stellar masses, their separation, and the relative speed of the wind. In what follows we show under which conditions accretion onto a star orbiting around  a massive companion can be drastically  affected by the presence of the companion.

\subsection{Analytical Derivation}
\label{sec: analytical}
We set up the problem following the Bondi-Hoyle-Lyttleton (BHL) \citep{Bondi1944, hoyle_lyttleton_1939} formalism, where we have the supersonic motion of a massive gravitational particle in a uniform density medium. The gas around the particle gravitationally focuses behind it, allowing it to accrete a substantial amount of material. We characterize the gas by a density $\rho_{\infty}$ and a Mach number $\mathcal{M} = v_{\infty} / c_{\rm{s},\infty}$ where $c_{\rm{s},\infty}$ is the speed of sound in the medium and $v_{\infty}$ is the velocity of the massive particle with respect to the background medium. 

We start with a massive particle of initial mass $M$ to obtain an accretion radius, 
\begin{equation}
    R_a = \frac{2GM}{v_{\infty}^2},
\end{equation}
which gives us a length scale for the gravitational influence of the particle on the background gas. By setting this length scale, we can calculate the Bondi mass accretion rate as the flux of gas through a circular cross section with radius $R_a$, that is, 
\begin{equation}
    \dot{M_{\rm B}} = \pi R_a^2 \rho_{\infty} v_{\infty} 
    = 4 \pi G^2 M^2 \rho_{\infty} v_{\infty}^{-3}.
\end{equation}

For a fixed mass $M$ and velocity $v_{\infty}$, we note that $\dot{M}_{\rm B} \propto \rho_{\infty}$. For a full review of the Bondi-Hoyle-Lyttleton accretion formalism we refer the reader to \cite{Edgar_2004}. For our case of interest, we generalize the BHL formalism to encompass a binary system comprised of a black hole of mass $M_{\rm{BH}}$ with a companion star of mass $M_{*}$. A key goal of this paper is to understand how $M_{*}$ is affected by the nearby gravitational potential of the black hole. We take the cases where the separation $a \lesssim R_a$ such that the black hole and its companion share the same shock structure. We define a mass ratio between the black hole and companion star as $q = M_*/M_{\rm{BH}}$. The accretion radius of the binary becomes, 
\begin{equation}
    R_{a, \rm{bin}} = \frac{2G(M_{\rm{BH}}+ M_*)}{v_{\infty, \rm{CoM}}^2}
     = \frac{2GM_{\rm{BH}}(1+q)}{v_{\infty, \rm{CoM}}^2},
\end{equation}
where ${v_{\infty, \rm{CoM}}}$ is the relative velocity of the center of mass of the system. As the mass ratio decreases ($q \ll 1$), we expect the accretion radius to become, 
\begin{equation}
    R_{a, \rm{bin}} \approx \frac{2GM_{\rm{BH}}}{v_{\infty, \rm{BH}}^2} = R_{a, \rm{BH}},
\end{equation}
thus approaching the accretion radius of the black hole. We note that the velocity that dominates for this case is the relative motion of the black hole with respect to the background material, since it is much larger that its  orbital velocity around the center of mass. That is, $v_{\infty,\rm{BH}} \approx v_{\infty} \gg v_{\rm{orb},\rm{BH}}$. As for the companion star, we expect the accretion radius to be,
\begin{equation}
    R_{a,*} = \frac{2GM_*}{v^2_{*}},
    \label{eq: Ra_star}
\end{equation}
where $v_{*}$ is the velocity of the companion star with respect to the background gas. There are two components that contribute to $v_*$, the center of mass velocity of the system, $v_{\infty}$, and the orbital velocity of the star around the center of mass of the binary, $v_{\rm{orb}, *}$. The largest of these two velocities will be the dominant term in Eq. \ref{eq: Ra_star}. That is, if $v_{\infty} \lesssim \sqrt{GM_{\rm{BH}}/a}$, the orbital velocity dominates. The two velocities are equivalent at a separation of $a = 0.5 R_a$. If we start with the assumption that $v_* = v_{\infty}$, valid for $0.5 R_a \lesssim a \lesssim 1.0 R_a$, we can calculate the mass accretion onto the companion star, which is the flux of material within one accretion radius $R_{a,*}$, for a star in isolation, as 
\begin{equation}
    \dot{M}_{*, \rm{iso}} = \pi R_{a,*}^2 \rho_{\infty} v_{\infty}. 
\end{equation}

The main distinction that occurs once the system is in a binary, instead of in isolation, is the increase in background density being sampled by the star. The potential created by the black hole will significantly enhance the concentration of material in its vicinity. We expect the relative density, $\rho_{\rm{rel}}$, to depend on the mass of the main driver of the potential, i.e. the massive black hole companion, and on the overall separation of the star from the cusp of the density concentration profile. Given these conditions, we expect the relative density to be of the form, 
\begin{equation}
    \rho_{\rm{rel}} = \rho_{\infty} 
    \left(
    \frac{M_{\rm{BH}}}{M_*}
    \right)^{\xi} 
    \left( \frac{a}{R_a} \right)^{-\beta},
    \label{eq: rho_rel}
\end{equation}
where $\xi$ and $\beta$ are parameters left to be fit by the numerical simulations. Using the relative density now, the mass accretion rate of a star in a binary becomes, 
\begin{equation}
    \dot{M}_{*, \rm{bin}} = \pi R_{a,*}^2 \rho_{\rm{rel}} v_{\infty}
    = 4 \pi G^2 M_*^2 \rho_{\rm{rel}} v_{\infty}^{-3}. 
    \label{eq: mdot}
\end{equation}

We now get a relation between the accretion rate of the star in isolation and the accretion rate of the star in a binary, 
\begin{equation}
    \dot{M}_{*, \rm{bin}} =
    \dot{M}_{*, \rm{iso}}
    \left(
    \frac{M_{\rm{BH}}}{M_*} 
    \right)^{\xi} 
    \left( \frac{a}{R_a} \right)^{-\beta}.
    \label{eq: xi_beta}
\end{equation}

The values of $\xi$ and $\beta$ are not evident from the analytical approach, which leads us to find numerical approximations for them. We do this by holding one of the values constant while varying the other parameter. We find that, within numerical inaccuracy, $\xi \approx \beta \approx  1$. We refer the reader to Section~\ref{sec: dependencies} for the justification. Using these values, we calculate
\begin{equation}
    \dot{M}_{*, \rm{bin}} 
    = \left(
    \frac{M_{\rm{BH}}}{M_*}
    \right)
    \left( 
    \frac{a}{R_a} 
    \right)^{-1} 
    \dot{M}_{*, \rm{iso}}.
\label{eq: m_dot_bin}
\end{equation}
For a star orbiting a black hole with $M_{\rm BH}=100\,M_\ast$ and $R_a/a=0.1$, for example, this implies $\dot{M}_{*, \rm{bin}}=10^3\dot{M}_{*, \rm{iso}}$. We note that one can do a similar calculation for the case where the dominant velocity is $v_{*,\rm{orb}}$. This gives 
\begin{equation}
    \dot{M}_{*, \rm{bin}} 
    = \left(
    \frac{M_{\rm{BH}}}{M_*}
    \right)
    \left( 
    \frac{a}{R_a} 
    \right)^{-\frac{1}{2}} 
    \dot{M}_{*, \rm{iso}}.
\end{equation}
The only difference in dependency that comes into play is the separation, where $\beta = \frac{1}{2}$ instead of 1. The analytical solution indicates that for separations $a \lesssim 0.5$, the orbital solution is expected to dominate. This assumes the star is moving through unperturbed gas, that is unaffected by the binary's presence. Instead, the star is embedded in a dense background that has already been gravitationally focused and shocked by the massive black hole. When we look closely at the relative orbital velocity of the star with respect to the gas, we see that it is always lower than $v_{\infty}$ for significantly lower radii than the analytical solution predicts. We show  this result in the Appendix~\ref{app:velocity}. We therefore conclude that using the relation in Eq.~\ref{eq: m_dot_bin} holds true for the majority of close binaries. 

\begin{figure*}[ht]
\centering
\includegraphics[width=\textwidth]{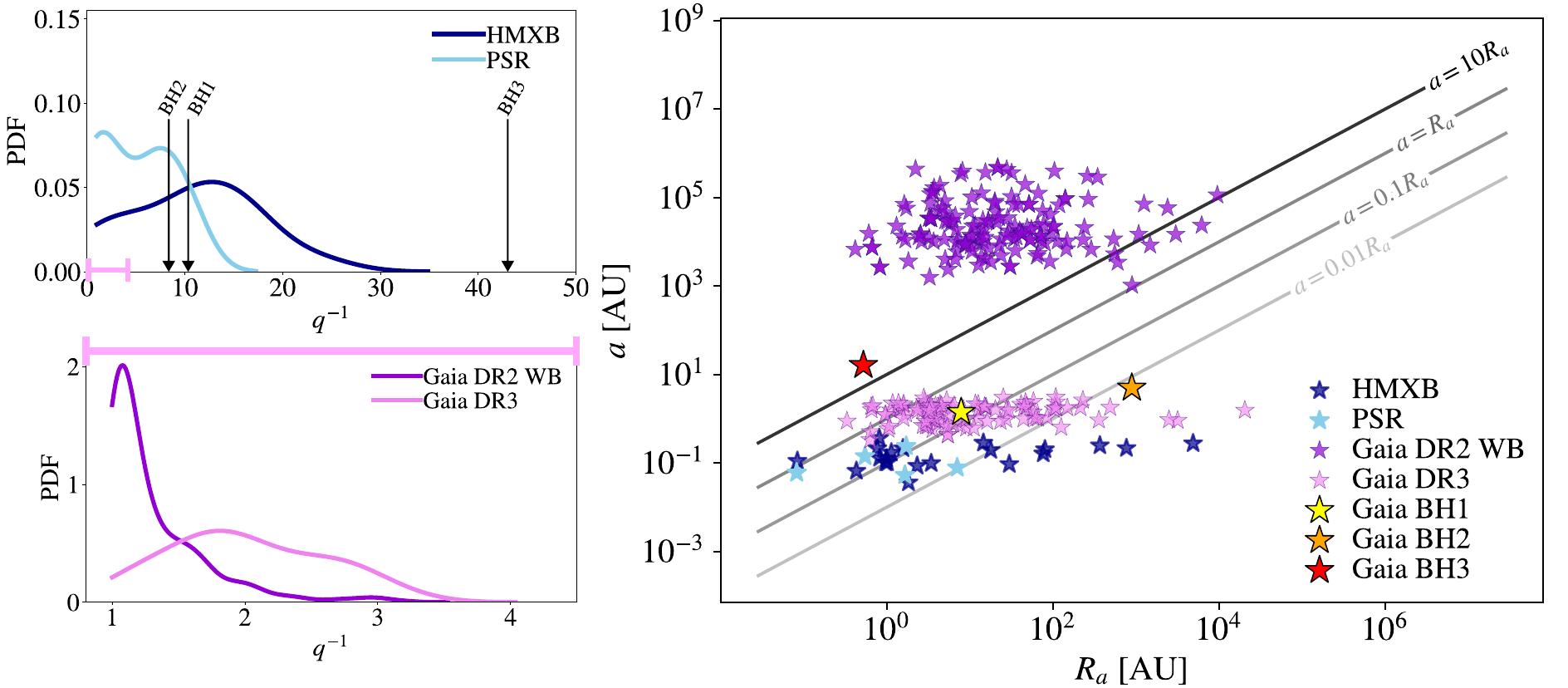}
\caption{\textit{Left:} Probability distribution of the mass ratio of binaries in three different catalogs: \textit{top:}  HMXBs (dark blue), radio pulsars (light blue), and \textit{bottom:}  \textit{Gaia} wide binaries (purple) and \textit{Gaia} DR3 binaries (pink). The locations of the three recently discovered \textit{Gaia} black holes are labeled by the three black lines in the top pannel.
\textit{Right:} Semi-major axis distance between two orbiting bodies, $a$, versus accretion radius, $R_a$ for different detected binary systems. The diagonal lines shows the location where $a/R_a = 10, 1, 0.1, \ \text{and} \ 0.01$ from top to bottom (darker to lighter). In the upper left, denoted by purple stars, is a subset of the wide binaries detected by \textit{Gaia} DR2. Lower down, in pink, is a subset of the binaries detected by \textit{Gaia} in the DR3 non-single stars catalogue. In the bottom, in dark blue are the HMXBs and in light blue the radio pulsar binaries. The stars in yellow, orange and red, outlined in black, show the three \textit{Gaia} black hole binaries.}
\label{fig: a_vs_ra}
\end{figure*}

\subsection{The Landscape of Unequal Mass Binaries}
\label{sec:landscape}
In an effort to understand the role that massive companions might have on the accretion history of lighter binary members, we explore here the  mass ratio distributions in binaries. While the average mass ratio for solar-type binaries is $q \approx 1/2$ \citep{Moe2019}, we expect $q$ to go from equal mass, $q=1.0$, to extreme, with $q \lesssim 10^{-4}$, as in the case of extreme mass ratio inspirals around supermassive black holes \citep{2007CQGra..24R.113A}. Apart from $q$, the binary separations will also range vastly from compact binaries with separations on the order of tens of $R_{\odot}$ to wide binaries with separations larger than millions of AU. To illustrate the diversity of binaries, we plot a selection of representative examples in Figure~\ref{fig: a_vs_ra}.

Figure~\ref{fig: a_vs_ra} shows an illustrative portion of the observed binaries in our local Universe selected  from three catalogs: \textit{Gaia} DR2 wide binaries (WBs) \citep{gaia_wide_binaries}, \textit{Gaia} DR3 non-single stars \citep{2023A&A...674A..34G}, HMXBs \citep{Bildsten_1997, Tsygankov_2022} and radio pulsar binary systems \citep{Rivinius_1997, Clark_2000}, which we denoted as PSR. In the right panel of Figure~\ref{fig: a_vs_ra}, we compare the two key length scales of the problem (i.e., $a$ and $R_a$). The $y$-axis shows $a$ while the $x$-axis shows $R_a$ as defined in Section~\ref{sec: analytical}, both in AU. The diagonal lines represents fixed separation to the accretion radius values, such that $a/R_a = 10, 1, 0.1, \ \text{and} \ 0.01$. Any binaries with $a/R_a \gtrsim 1$  will have two independent accretion radii, while any binaries below the $a/R_a = 1$ line will share an accretion structure. The purple stars at the top left of the plot are \textit{Gaia} DR2 wide binaries, which are observationally biased towards systems with large separations. Below them, in pink, is a subset of the \textit{Gaia} DR3 binaries. {These stars were selected by a query similar to that of \cite{2023AJ....166....6C}, with the main difference of lowering the restrictions for $\texttt{m2}$ (lower confidence level of the secondary mass) and $\texttt{rv\_nb\_transits}$ (number of transits used to calculate the radial velocity) to be $\texttt{m2\_lower > 1}$ and $\texttt{rv\_nb\_transits > 5}$. In dark (light) blue are HMXBs (PSR), which have all shared accretion structures. }The yellow, orange and red points are the \textit{Gaia} black hole systems \citep{2023MNRAS.518.1057E,2023MNRAS.521.4323E,2023AJ....166....6C,gaia_bh3}, denoted as BH1, BH2 and BH3, respectively. As we show in  Section~\ref{sec:dis_ms_bh3} with the use of cosmological simulations, we expect the relative velocity of BH3 with respect to the gas to be significantly smaller than its current velocity and, as a result, BH3 is expected to have had a shared accretion structure for the majority of its enhanced mass accretion history.

In the left panels of Figure~\ref{fig: a_vs_ra}, we choose to depict the probability distribution of mass ratios for the four different sets of binary stars. We take the mass ratio $q=M_2/M_1$ where $M_1$ is the mass of the more massive star, which we refer to as the primary, and $M_2$ as the less massive star, which we refer to as the secondary. The choice of plotting $q^{-1}$ is done to make it easier to observe the wide range of mass ratios in the data. For the \textit{Gaia} DR2 WBs in the bottom panel, there is an early peak which shows a strong tendency towards similar mass components. For the \textit{Gaia} DR3 case, the peak is less pronounced but there is an overall preference towards similar inverse mass rations. In the top panel, we plot the radio pulsars which have two  distinct peaks, while the HMXBs show most unequal mass ratio distribution, with a peak around 15. We see that the \textit{Gaia} black holes BH1 and BH2 have $q^{-1}$ of about 10, while \textit{Gaia} BH3 is towards the extreme tail end of the distribution, with $q^{-1} \approx 42$. The splitting of the populations between two panels enables a clearer depiction of the overall distributions, as the range of mass ratios of both of the \textit{Gaia} catalogs is much smaller than that of the other two populations and the three BHs. We use these mass ratio distributions to inform our hydrodynamic simulations and, in particular, we explore the chemical enrichment of metal-poor stars in BH3-like systems  via enhanced accretion.

\begin{figure*}[ht]
    \centering
    \includegraphics[width=0.98\linewidth]{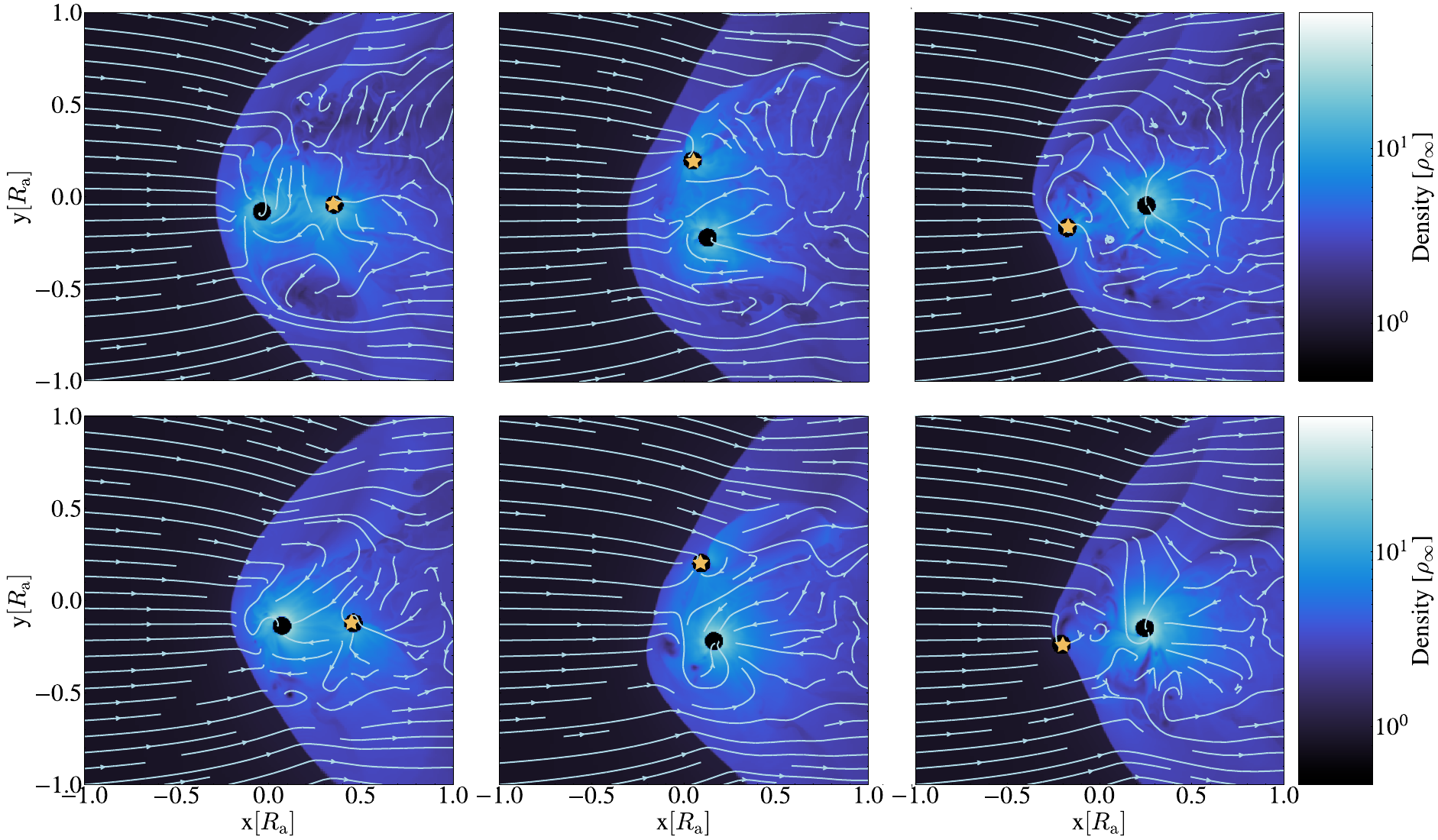}
    \caption{Density snapshots for two simulations at three different time stamps ($t=55, 61, 67 \ R_a/v_{\infty}$) as the system completes half of an orbit from left to right. Higher densities are denoted by lighter colors. The stellar companion is represented by the yellow star marker. The arrowed flow lines  show the velocity streamlines of the incoming wind. \textit{Top row:} Time evolution of a system with mass ratio $q^{-1}=2$ at a separation $a = 0.42\,R_a$. 
    \textit{Bottom row}: Time evolution of a system with mass ratio $q^{-1}={5}$ at a separation $a = 0.42\,R_a$.
    }
    \label{fig:snapshots}
\end{figure*}

\section{Methods and Numerical Setup} \label{sec: hydro}
To simulate the accretion onto a lighter companion, we set up an unequal mass binary in a wind tunnel. In this setup, the black hole-star binary is originally placed at rest at the origin of the grid, while a uniform wind is injected from the left boundary. Following \citet{Antoni2019}, the binary's orbit is initially circular and is allowed to evolve self-consistently  as it accretes material (and the corresponding momentum) from the background gas and is subjected to drag forces. The center of mass of the system is not fixed at the origin, so the reaction of the binary to the gaseous medium may induce a center of mass motion. 

We use the FLASH hydrodynamics code \citep{Fryxell2000} to solve the fluid equations over time. FLASH is a Eulerian code and we use the directionally split Piecewise Parabolic Method Riemann solver \citep{Colella_1984}.  We construct a three-dimensional hydrodynamics setup where a uniform wind is injected with velocity $v_{\infty}$ from the $-x$ boundary in the $+x$ direction. The $+x$, $\pm y$, and $\pm z$ boundaries are all outbound boundaries, while the $-x$ is an inbound boundary.  The computational domain is $(-3, 3)R_a \times (-3, 3)R_a \times (-3, 3)R_a$. The maximum and minimum refinement levels are set to 6 and 2, respectively, with a maximum cell size of $R_a/16$ and a minimum of $R_a/256$. 

The black hole-star binary is modeled by adding two active sinks \citep{Federrath_2010}, which allows us to track the accretion and forces on both the black hole and the companion throughout the simulation. We initially place the center of mass of the system at the origin of the grid, $R_{\rm{CoM}} = 0$ at time $t=0$. The center of mass of the binary is set to be at rest, $\bf{v}_{\rm{CoM}} = 0 $, and the particles are put in a circular orbit in the $x-y$ plane with separation $a$ and angular momentum vector pointing in the $+z$ direction. The sinks have a physical radius of $R_s = 0.05R_a$, based on  resolution studies, assuring that $R_s/\delta_{\rm{min}} = 3$, where $\delta_{\rm{min}}$ is the length of the finest cell size \citep{Antoni2019}. The total mass of the system is fixed from one simulation to another, such that $M_{\rm{BH}} + M_* = M_{\rm{T}}$ where $M_{\rm{T}}$ is set to $(2G)^{-1}$ and $q={M_*}/{M_{\rm{BH}}}$. The simulations are carried out in dimensionless units where we set $v_{\infty} = 1$ and $\rho_{\infty} = 10$.  This yields an accretion radius of $R_a = 1$ which we use as the length scale of the problem.  Time in the simulation has units of $R_a/v_{\infty}=1$. 

We investigate mass ratios of \(q^{-1} = 2, 3, 4, 5\) for two distinct separations: \(a = 0.42\) and \(1.0 R_a\). Simulations with \(a < 0.42\) required significantly higher resolution and smaller sink radii to be reliably resolved \citep{Antoni2019}.
For most of the simulations, we investigate cases where the gaseous medium has an adiabatic index of $\gamma = 5/3$, except for the case where we keep $q$ and $a$ constant and vary $\gamma$. A $\gamma = 1.1$ equation of state is used to simulate the properties of the flow when significant cooling occurs in the gas, which we denote as the quasi-isothermal case. The  simulations with varying adiabatic index are presented in Appendix~\ref{app:gamma}. For all simulations in this work, the Mach number of the gas relative to the center of mass of the system is initially set to $\mathcal{M} = 2.0$. The use of dimensionless parameters allows the results from these simulations to be scalable to a variety of mass ratio binaries across a wide range of astrophysical regimes, especially for mass ratios that are difficult to resolve numerically ($q^{-1} \gtrsim 5$).

\section{Numerical Results}\label{sec: results}
In this section, we present the results of our numerical simulations. We analyze the mass accretion rate across the sink representing the less massive companion as well as the density structures that develop around the binary system. This analysis allows us to determine the average local density surrounding the less massive stellar companion, which in turn helps us estimate its mass accretion rate. By following this procedure, we can investigate how the mass accretion rate of the less massive companion varies with the mass ratio and separation of the binary.

\begin{figure*}[ht]
    \centering
    \includegraphics[width=0.99\linewidth]{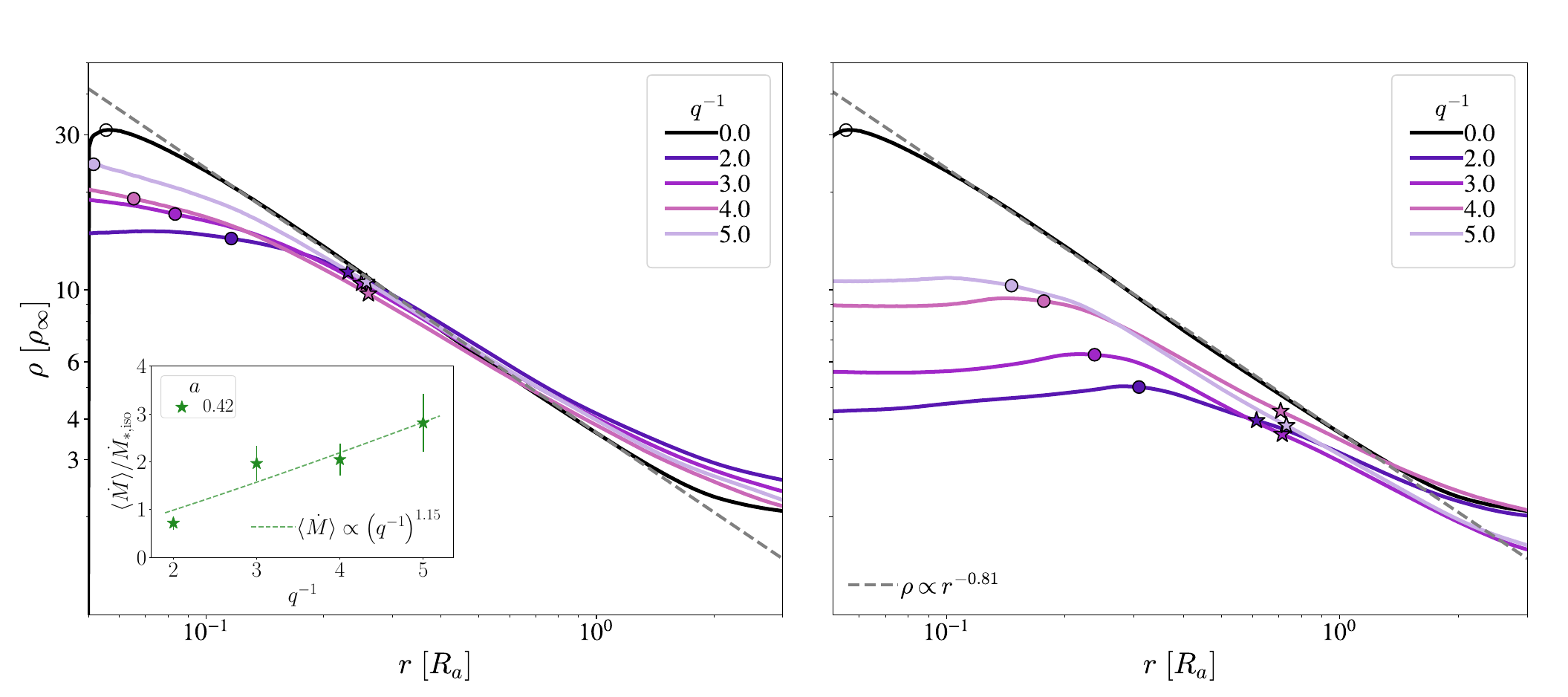}
    \caption{Density profiles from the center of mass of BH-star binaries for different mass ratios at a fixed initial separation of $a =0.42\,R_a$ (\textit{left}) and $a =1.0\,R_a$ (\textit{right}). From top to bottom, $q^{-1}=2,3,4,5$. The circles show the locations of the BHs and the star shaped symbols show the locations of the stars relative to the center of mass of the binary. We can see that the stars experience  densities that are $\approx 6-10$ times larger than the background one. The gray dotted line shows the best fit to the single object between $0.1 \text{ and } 1.0 R_a$, $\rho \propto r^{-0.81}$. \textit{Left inset}: Orbit averaged accretion rate for the stellar companion compared to its mass accretion rate in isolation. This is plotted for different inverse mass ratios $q^{-1}$, at fixed orbital separation of $a = 0.42$. The best fit line is plotted which yields the relation $\langle  \dot{M} \rangle \propto \left( q^{-1} \right) ^{1.15\pm 0.2}$. 
    }
    \label{fig: dens-redist}
\end{figure*}

\subsection{Flow morphology}
We illustrate here the flow morphology of the shared shock structure between the black hole and its stellar companion. The snapshots presented in Figure~\ref{fig:snapshots} are taken in the orbital plane ($z = 0$), and the colormap illustrates the density of the flow, with lighter colors denoting higher densities. Flow lines are drawn to show the velocity streamlines of the material around the binary. The snapshots are taken at three different times for half an orbit (left to right). The stellar companion is labeled with the yellow star marker and the black hole is labeled with the black circle. The top three panels show the time evolution of a binary with a mass ratio of $q^{-1}=2$ and an orbital separation of $a=0.42 R_a$. The bottom three panels show the time evolution of a binary with a mass ratio of $q^{-1}=5$ and an orbital separation of $a=0.42 R_a$.

We note that for the larger mass ratio case (\textit{bottom} panels in Figure~\ref{fig:snapshots}), there is a higher concentration of material near the black hole, illustrated by the enhanced light blue region. The shock is also more pronounced as the stellar companion has less influence on the overall potential. For the lower mass ratio case (\textit{top} panels in Figure~\ref{fig:snapshots}), the potential is more closely affected by both components of the binary, slightly widening the shock structure and pushing forward more shocked material when the star crosses in front (upstream) of the black hole. This can be seen by the separation of the streamlines going directly into the star, as opposed to into the black hole, in the right most panels in  Figure~\ref{fig:snapshots}. It is clear that for a given mass ratio, as the orbital separation of the system decreases, we expect the companion to be able to encounter higher density material and, as such, accrete gas at higher rates.

\subsection{Mass ratio and separation dependencies}\label{sec: dependencies}
We investigate how the local density experienced by the less massive companion in binary systems changes with separation for different mass ratios. Figure~\ref{fig: dens-redist} shows the density distribution for two different initial separations $a/R_a$, each featuring four distinct mass ratios: $q^{-1} =2,3,4,5$. The uppermost black line labeled as $q^{-1} = 0$ denotes the density profile for the case of a single object with total mass equal to that of the binary. 

For both of the initial separations studied in Figure~\ref{fig: dens-redist}, the density redistribution approaches that of the single object as the mass ratio decreases ($q^{-1}$ increases). That is, for highly unequal mass ratios, the density redistribution profile will closely resemble the profile of a single mass. This is to be expected, as the smaller object has a diminishing  influence and the majority of the redistribution of the gas is governed by the massive black hole. In these examples, where $a/R_a$ is held constant, we can see the trend where the physical separation between the objects increases as $q$ decreases.
This trend is more pronounced in cases where the initial ratio \( a/R_a \) is smaller, such as \( 0.42 \) compared to \( 1.0 \). In the single body scenario, the peak density reaches approximately 30 times the background density. In contrast, the secondary stars experience around 6 times the density enhancement for \( a/R_a \approx 1.0 \) and about 10 times the enhancement for \( a/R_a \approx 0.42 \).

The overall trend for each separation is that the mass distribution approaches that of a single point mass as the mass ratio decreases (the masses of the two objects are highly unequal). From this trend, we  extrapolate that the change in the relative density, as presented in Eq~\ref{eq: rho_rel}, evolves as $\rho_{\rm{rel}} \propto  r^{ -\beta}$, with $\beta \approx 1\pm 0.2$. In what follows we approximate the value of $\beta$ to be $\approx 1$.  

In the left inset of Figure~\ref{fig: dens-redist}, we plot the orbit averaged mass accretion rate as a function of $q^{-1}$. The mass accretion rate values are normalized to the mass accretion rate of the star if it was found in isolation in a medium with background density $\rho_{\infty}$. We derive $\langle \dot{M} \rangle \propto \left(q^{-1} \right)^{\xi} \propto \left(q^{-1} \right) ^{1.15\pm 0.2}$. The best fit relation for $a = 1.0$ is also consistent with $\xi\approx 1$. Here we directly probe the relation between the mass ratio and mass accretion rate presented in Eq.~\ref{eq: xi_beta}. We conclude that the density enhancement experienced by the companion is altered by the mass ratio as $\rho_{\rm{rel}} \propto (q^{-1})^{1.15\pm 0.2} \approx q^{-1}$.  In high-density environments, cooling will further enhance the mass accretion rate (Appendix~\ref{app:gamma}). Thus, these relationships should be seen as offering a conservative limit.

\section{Chemical enrichment of metal-poor stars orbiting \textit{Gaia} BH3-like Black Holes} \label{sec: case study}

A recently discovered black hole orbiting a low-mass star was detected using astrometry with the \textit{Gaia} satellite. This system consists of a black hole with a mass of \(32.70 \pm 0.82\,M_{\odot}\), which is orbited by a star with a mass of \(0.76 \pm 0.05\,M_{\odot}\) and a measured metallicity of \([{\rm Fe/H}] = -2.56 \pm 0.1\). The system has an orbital period of \(4,235.1 \pm 98.5\) days, a semi-major axis of \(16.17 \pm 0.27 \, \text{AU}\), and an eccentricity of \(0.7291 \pm 0.0048\) \citep{gaia_bh3}. This system acts as a test bed for studying the accretion history of low metallicity stars that orbit massive black hole companions.

We propose that the ISM pollutes low-mass companions through Bondi accretion. The ability of these companion stars to capture newly synthesized metals from the ISM is significantly enhanced by the gravity of their massive partners. In this scenario, the surface compositions of these companions would differ from their cores, with low-mass stars that have the thinnest convective layers experiencing the highest degree of pollution (i.e., less dilution). Low-mass stars, such as the companion of BH3, undergo mixing  as they evolve away from the main sequence \citep{2014PASA...31...30K}. The first mixing event is known as the first dredge-up, which occurs once the hydrogen in the core is depleted. As the core contracts, it leads to the expansion of the star and the deepening of its convective envelope. The first dredge-up submerges and dilutes any species that are only present on the stellar surface \citep{2000A&A...354..169G,2020A&A...638A..94O}. This occurs, for instance, when a star accretes material from a disrupted planet \citep[e.g.,][]{2018ApJ...853L...1M,2021AJ....162..273S,2023ApJ...954..176Y}. The resulting signatures in the star's photosphere are influenced by internal mixing processes, which can weaken or alter the observed enrichment over time \citep{2021AJ....162..273S}.

To quantify the observational signature of external pollution, we use, in what follows,  Modules for Experimentation in Stellar Astrophysics \citep[MESA;][]{Paxton2011,Paxton2013,Paxton2015,Paxton2018,Paxton2019,Jermyn2023} and adopt the \texttt{low\_z} test suite inlist. This setup facilitates the evolution of $0.8\,M_{\odot}$ low metallicity stars from the pre-main sequence through the asymptotic giant branch phase. The range of masses of the convective region during the main sequence is $1.47 \times 10 ^{-3}\,M_{\odot}$ for stars with initial [Fe/H]~$= -2$ and $1.07 \times 10 ^{-3}\,M_{\odot}$ for [Fe/H]~$= -5$.
If the ISM contaminated the surface of the star during its main sequence phase, convection would have mixed this material throughout the convective envelope. When the star transitions off the main sequence, the first dredge-up would naturally decrease the surface abundance of the accreted elements \citep[e.g.,][]{2020ApJ...891L..13K}. 
Inspired by this, we examine the accretion history of low metallicity main-sequence stars orbiting massive black hole companions (Sections~\ref{sec:dis_ms_bh3} and \ref{sec:dis_ms}) and analyze how the resulting pollution signatures change as the companion stars evolve (Section~\ref{sec:dis_bh3}). If this scenario is occurring in \textit{Gaia} BH3, we would expect a depletion in the abundances of accreted metals at the surface of the companion star.

\subsection{The mass accretion history of \textit{Gaia} BH3-like binaries in the Milky Way}\label{sec:dis_ms_bh3} 
In order to calculate the amount of metals accreted over the companion star's history, we use the stellar trajectories derived by \citet{Shen_2017} from the \textit{Eris} simulation \citep{Guedes_2011}. These trajectories have information about the velocity of the system as well as the density and metallicity of the gas encountered by the stellar particles in the simulation. The \textit{Eris} simulation is a high-resolution cosmological zoom-in simulation of a Milky Way analog from redshift $z=12$ to $z=0$. In the simulation, new star particles inherit the metallicity of their parent gas particles. These star particles return energy, mass, and metals to the ISM through core-collapse and type Ia supernovae \citep{2015ApJ...807..115S}. For further details about the \textit{Eris} simulation, we refer the reader to \citet{Guedes_2011}, and for details on the stellar particle tracing we refer the reader to \citet{Shen_2017}. 

Following \citet{Shen_2017}, we calculated the total amount of Fe accreted onto the $0.8\,M_{\odot}$ companion star over its evolutionary history. This star is  orbiting  a $32.7\,M_{\odot}$ black hole at a separation of 16\,AU mimicking the properties of \textit{Gaia} BH3. Given its Galactic coordinates, Gaia BH3 is classified as stellar system belonging to the disk in the \textit{Eris} simulation \citep{Pillepich_2015}, where approximately 77\% of stars formed in situ (i.e., within the Milky Way). In-situ stars are expected to experience higher accretion rates over their history, as the Galactic star-forming environment is typically denser and more metal-rich than for stars residing in satellite galaxies.

Using the fits for $\xi$ and $\beta$ in Equation~\ref{eq: m_dot_bin}, we calculate the overall factor by which the relative density experienced by the star is augmented due to its black hole companion. That is, $\rho_{\rm{rel}} =\rho_{\infty} (M_{\rm{BH}}/M_*) (R_a/a)\approx 40 \rho_{\infty} (R_a/a)$. This allows us to calculate the accretion rate onto the low-mass orbiting companion for each of the particle trajectories outlined in \citet{Shen_2017}, with the relative velocity to the gas being calculated self-consistently. We utilize a uniform random sample of 2,000 metallicity-selected star particles, which were selected by \citet{Shen_2017} from all the {\it Eris} star particles residing in the halo, to compute the metal accretion.

\begin{figure}[ht]
    \centering
    \includegraphics[width=0.95\linewidth]{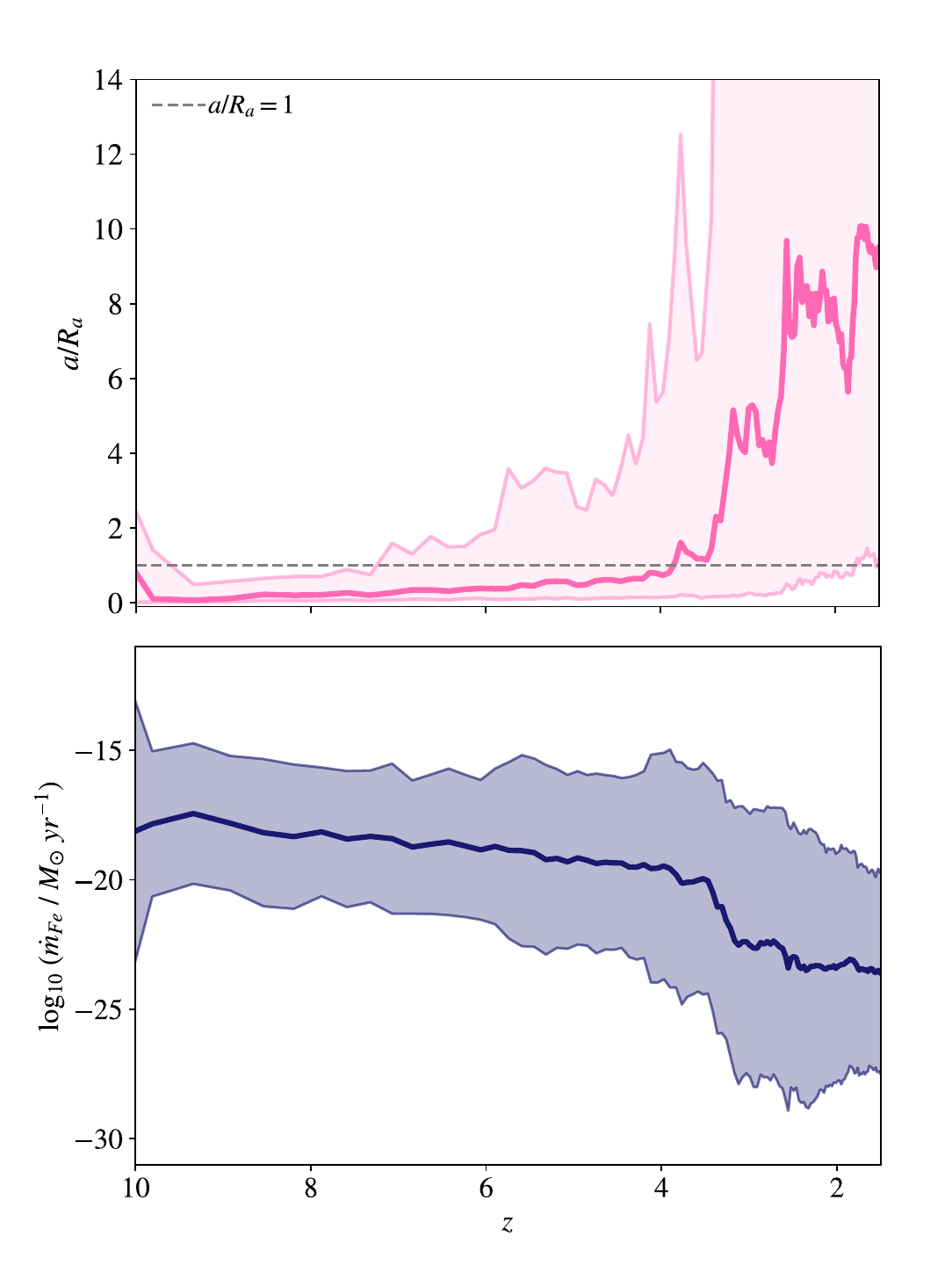}
    \caption{\textit{Top:} Evolution of $a/R_a$ for a \textit{Gaia} BH3-like  binary. Namely, a $0.8\,M_{\odot}$ companion star orbiting around  a $32.7\,M_{\odot}$ black hole at a separation of $16\,\rm{AU}$. The components used from \textit{Eris} simulations are the  velocity of the star particles relative to the ISM, which determines $R_a$, as well as the gas density as a function of $z$. The gray dotted line denotes where  $a/R_a=1$, such that for all values under the line, the binary has a shared accretion radius.  \textit{Bottom:} Distribution of the companion’s cumulative iron accretion history from formation to the present day, derived from the same set of trajectories and accounting for the spread in relative velocities. In both panels, dark curves show the median values and the shaded region show the $68$\% scatter about the median. }
    \label{fig:redshift}
\end{figure}

{We calculate the evolution of $a/R_a$ and the corresponding iron accretion onto the low-mass companion in several steps. 
First, we compute $a/R_a$ for each stellar particle using its relative velocity. In the top panel of Figure~\ref{fig:redshift}, we show the evolution of $a/R_a$ for the binary system as a function of redshift $z$. Each binary experiences varying background densities over its evolutionary history. We use the local density, $\rho_\infty$, together with $a/R_a$ to scale the accretion rate for each individual trajectory.
Second, we integrate the accreted iron along each trajectory to obtain the mass accretion history of the companion. The bottom panel of Figure~\ref{fig:redshift} shows the resulting iron accretion rate as a function of redshift. The majority of the accretion occurs at $z\gtrsim 4$, with  a sudden drop after this. This drop is associated with the last major merger in the galaxy \citep{Shen_2017}.}

To quantify the observational signature of external pollution, we integrate the total amount of iron accreted using values derived from the particle trajectories of halo stars \citep{Shen_2017} and assume an initial birth [Fe/H] of $-5$ to calculate the observed metallicity of the polluted surface of the stellar companion. For redshifts where $a/R_a < 1$, we apply a multiplicative factor to account for enhanced accretion, while for redshifts where $a/R_a \gtrsim 1$, we assume isolated accretion.  In most cases, the companion's contribution to the total metal mass accreted is negligible when $a/R_a \gtrsim 1$, due to the sharp decline in the accretion rate.

{There is a range of outcomes that can be obtained by integrating over all the particle trajectories. To illustrate how the results vary, we perform two calculations. In the first, we use the mean value of $a/R_a$ across all trajectories and scale the accretion rate using the mean background density, $\rho_\infty$. In this case, the metallicity of the polluted surface is [Fe/H] $= -3.4$. We also explore the impact of variations in the background density by considering cases where $\rho_\infty$ is shifted by $\pm 1\sigma$, which captures the typical spread in the environment experienced by the particles. With these variations, the metallicity of the polluted surface ranges from [Fe/H] $= -4.5$ to $-2.3$. We then repeat the same exercise, but instead of varying the background density, we use the mean background density of all particles and examine the impact of variations in $R_a/a$ on the results, which depend on the evolution of $R_a$ as a function of the binary's velocity. We find surface [Fe/H] values ranging from $-4.0$ to $-2.11$, with a median value of $-3.4$ for a stellar companion on the main sequence. }

{Given the current $Gaia$ BH3's Galactic coordinates, we track the evolution of representative stellar particles with similar properties at the end of the \textit{Eris} simulations (i.e., $z=0$). These systems allow us to estimate the surface metallicity of the stellar companion to be approximately [Fe/H] $= -2.2$, assuming it were still on the main sequence (Section~\ref{sec:dis_bh3}).}

This analysis indicates that the metallicity of non-evolved companion stars orbiting massive black holes may be significantly altered from their original birth values. We explore this idea further in Section~\ref{sec:dis_ms}. When a stellar companion evolves off the main sequence, as is the case for the \textit{Gaia} BH3 companion, the initial dredge-up process can substantially reduce any excess surface abundance produced by accreted metals. In Section~\ref{sec:dis_bh3}, we analyze the expected level of metal depletion in the surface abundance of the \textit{Gaia} BH3 companion and discuss whether it is possible to place meaningful constraints on its birth metallicity. {We note that our simulations do not include radiative feedback from the black hole onto the surrounding gas. While this effect is not expected to strongly alter the metallicity results presented here, we provide an estimate of its impact in Appendix~\ref{app:feedback}.}

\begin{figure}[ht]
\centering
\includegraphics[width=0.41\textwidth]{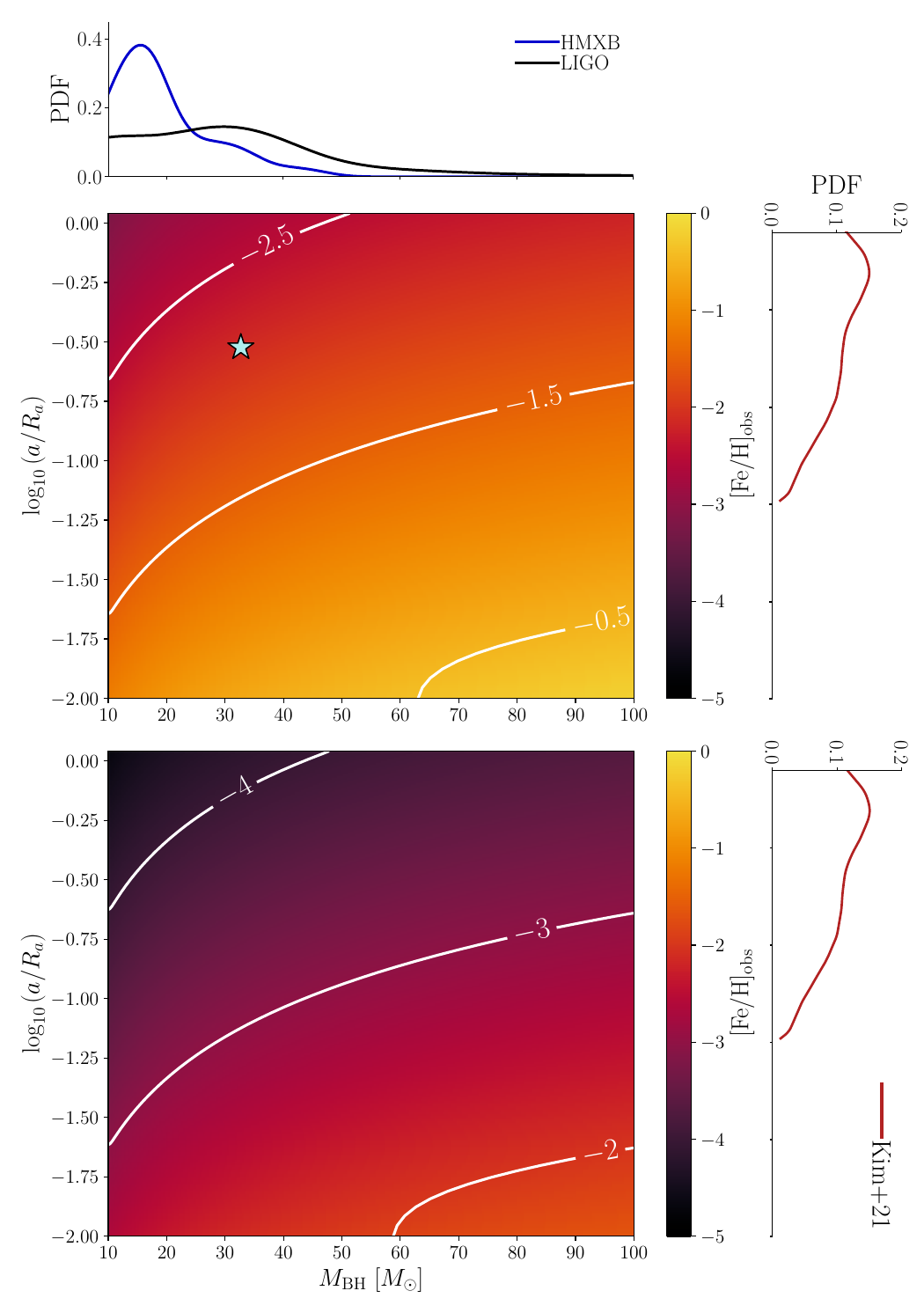}
\caption{{The surface abundance resulting from  accretion for a main-sequence star companion with $0.8\,M_{\odot}$, orbiting a $M_{\rm{BH}}$ black hole at a separation $a/R_a$. We assume the star has a birth [Fe/H]$=-5$.  The top panel shows the corresponding surface metallicity [Fe/H] calculated using the accretion history for  stellar particles with Galactic locations similar to  the \textit{Gaia} BH3  system. The teal star-shaped marker denotes the conditions for the $Gaia$ BH3  system if it were in the main sequence and when it was actively accreting in the early universe (see Figure~\ref{fig:redshift} where $a/R\approx 0.3$ for $z\lesssim 4$).  The bottom panel calculates the accretion history of the binary using the mean values of $a/R_a$  across all stellar particle trajectories and scaling the accretion rate with the mean background density, $\rho_\infty$ (Figure~\ref{fig:redshift}). For both cases, the surface abundance is progressively enhanced for stars residing at tighter separations with heavier black hole companions. 
\textit{Top panel distribution:} In black,  we show the probability density function of LIGO-Virgo-Kagra black hole masses \citep{LIGO}. In blue, we show the probability density function of the masses of the black hole component of low mass X-ray binaries \citep{Fortin_2024}. 
\textit{Right panel distribution:} We show the probability density function of [Fe/H] of \textit{Gaia} DR3 thick disc and halo stars \citep{Kim_2021}.
}}
\label{fig: a_m_z_plot}
\end{figure}

\subsection{Altering the birth metallicity of main-sequence companions to LIGO-like black holes}\label{sec:dis_ms}

{Here we relax the constraints derived from the specific configuration of the BH3 system  and expand our calculations to consider metal pollution in a $0.8\,M_{\odot}$  main-sequence star in a binary  system with varying separations ($a$ from $0.01$ to $1 R_a$) and  black hole  companion masses ($M_{\rm BH}$ from $10$ to  $100\,M_{\odot}$). We present the results from this exercise in Figure~\ref{fig: a_m_z_plot}, for which we have derived the total amount of iron accreted using the median value calculated from the particle trajectories of all halo stars  (bottom panel) and  \textit{Gaia} BH3--like stars (top panel) for redshifts where $a/R_a < 1$. We give, for simplicity, an initial birth [Fe/H] value of $-5$ to all stellar companions. }

This exercise allows us to probe a diverse collection of binary black hole-star systems that can inform future detections in the Milky Way. The range of stellar black hole masses in the Galaxy is closely linked to the life cycle of massive stars \citep[e.g.,][]{2010ApJ...725.1918O}, as well as the energy and dynamics associated with supernova explosions \citep[e.g.,][]{2016ApJ...821...38S,2018ApJ...862L...3S,2024PhRvL.132s1403V}. By studying this distribution through observations \citep[e.g.,][]{2020A&A...638A..94O}, we can gain insights into the types of black hole companions we might expect to find alongside low-mass stars.
In the distribution at the top of Figure~\ref{fig: a_m_z_plot}, we plot the mass distribution of the LIGO-Virgo-Kagra black holes \citep{LIGO} and the black hole components of low mass X-ray binaries \citep{Fortin_2024}. This distribution takes into account the masses of the primary and secondary components of the mergers, while black holes resulting from the merger were not considered. The low mass X-ray binary distribution peaks at around $15\,M_{\odot}$ while the LIGO distribution peaks at a larger value around $35\,M_{\odot}$. 

A few important points stand out from a close examination of Figure~\ref{fig: a_m_z_plot}. For binaries hosting  lighter black hole companions at large separations, the resulting surface abundances are low and closest to the star's birth metallicity. For binaries hosting high-mass black hole  companions at short separations, the surface abundance resulting from metal accretion can be significant. On the rightmost panel of Figure~\ref{fig: a_m_z_plot} we plot, for comparison, the metallicity distribution of the thick disc and halo stars from \textit{Gaia} DR3 \citep{Kim_2021}. This bimodal distribution has a peak near $-0.6$, which is attributed to the thick disk stars, and a peak around $-1.8$ attributed to the halo stars. We thus conclude that accretion onto low-mass, main-sequence companions, when enhanced by a massive black hole companion, can play a prominent role in altering their surface abundances. In these cases, we expect the inferred surface abundances to trend with stellar luminosity, as mixing processes become progressively more dominant as the star evolves.  

\begin{figure}[ht]
\centering
\includegraphics[width=0.5\textwidth]{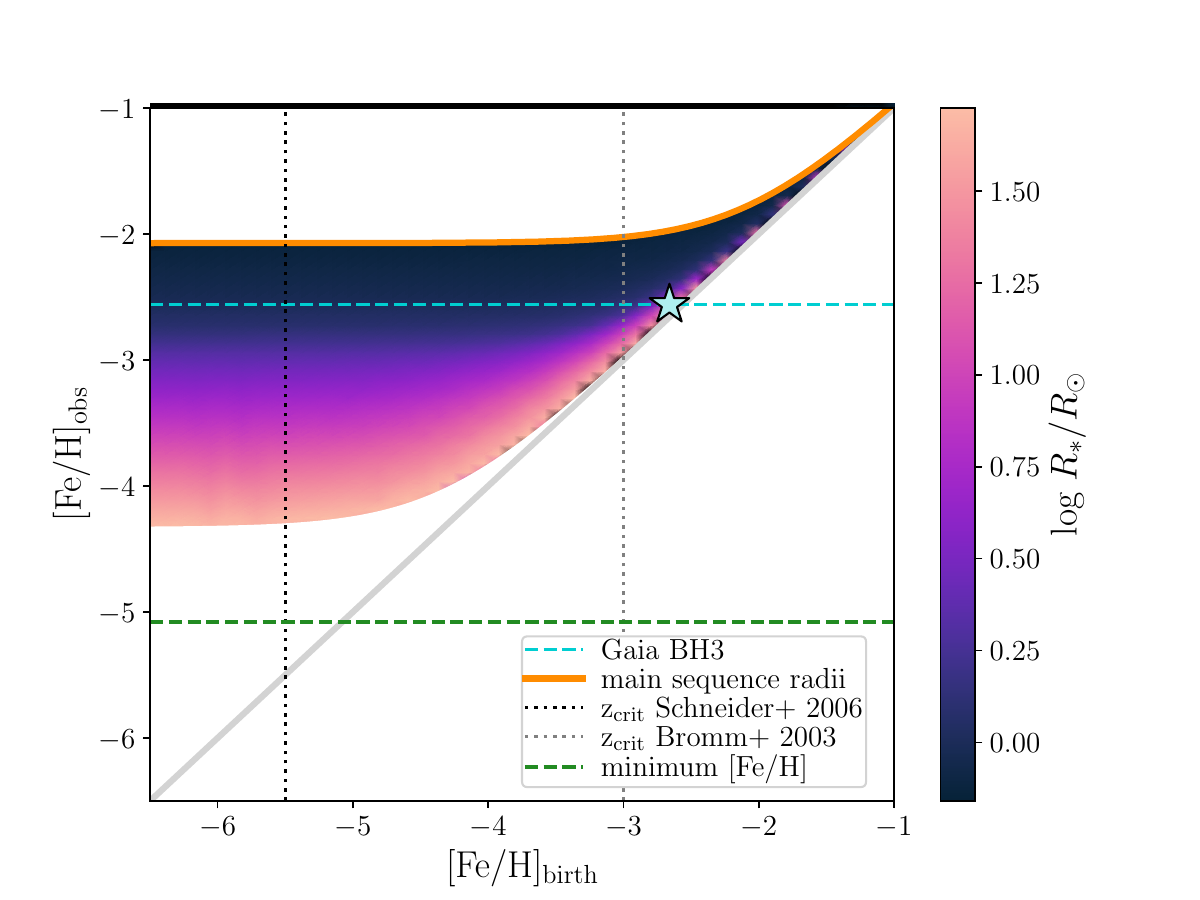}
\caption{The surface abundance enhancement via metal accretion of $0.8\,M_{\odot}$ companion stars with a given birth metallicity. The diagonal gray line shows the values for which the surface abundance is that of the star at birth. The color gradient indicates the different radii of the star as it evolves off the main sequence, with  darker colors  corresponding to more compact main sequence stars and lighter colors denoting extended giants. The orange curve denotes the surface abundance of stars in the main sequence. The horizontal teal dashed line shows the observed [Fe/H] value for \textit{Gaia} BH3, and the teal star marker matches the value of the observed metallicity to its current radii. The horizontal green line shows the minimum value for [Fe/H] if, at birth, the star was solely made up of hydrogen, all of its iron content came from accretion from the ISM and the polluting material was mixed into a fully convective star. The dotted vertical lines are critical metallicity  for population II  formation as derived by \citet{Bromm_2003} and \citet{Schneider_2006}.
}
\label{fig: z_plot}
\end{figure}
\

\subsection{On the birth metallicity of the \textit{Gaia} BH3 companion}\label{sec:dis_bh3}

We have demonstrated that the surface composition of the {\it Gaia} BH3 companion may have changed from its intrinsic composition due to metal pollution from the ISM, especially during its main sequence evolution. As the companion evolved beyond the main sequence, it would have experienced mixing as part of its evolutionary process \citep{2014PASA...31...30K}.  To quantify the observational signature of external pollution, we utilize MESA to calculate the mass of a star's convective region as a function of its age, which we characterize using its photospheric radius. This allows us to calculate the surface abundance of a 0.8\,\( M_{\odot} \) star throughout its evolutionary history for a range of initial birth metallicities. We plot these results in Figure~\ref{fig: z_plot}. 

The diagonal line in Figure~\ref{fig: z_plot} indicates when the surface abundance of the star is equal to its birth metallicity. The colored region denotes the parameter space in which the surface metallicity is higher than the birth metallicity due to accretion from the ISM material, as detailed in Figure~\ref{fig:redshift}. The darker colors in the diagram represent stars on the main sequence, which have shallower convective envelopes. In contrast, the lighter colors indicate stars in the giant branch, characterized by progressively larger convective envelopes. The orange curve denotes stars on the main sequence. The evolution of the star past the main sequence is then represented as a straight line downwards, expanding in size while decreasing in observed [Fe/H]. The horizontal teal dashed line represents the observed value of [Fe/H] for the \textit{Gaia} BH3 star and the star-shaped symbol correlates this value with the observed radius of the star. 

As a greater percentage of the star becomes convective, the metals accumulated over time are mixed more efficiently and become diluted. This dilution tends to diminish any signs of metal pollution, leading to the value of the observed metallicity to approach the value at birth. {As a comparison point, we calculate the observed metallicity of a fully convective $0.8\,M_{\odot}$ star made up of solely hydrogen that has been accreting from the wake of a $32.7\,M_{\odot}$ companion for its entire history. We argue that, although unrealistic, this scenario presents the limit  of the minimum observed metallicity given than a $100\%$ hydrogen star is the extreme limit of metallicity and a fully convective star is the case of maximal mixing of the accreted material. We plot the value obtained from this calculation as a green horizontal line in Figure~\ref{fig: z_plot}. } For guidance, we also provide the critical metallicity estimates for Population II star formation, $\rm{Z}_{\rm{crit}}$, as derived by \citet{Bromm_2003} and \citet{Schneider_2006}, represented here as the two vertical dotted lines.

The \textit{Gaia} BH3 system features a G-type giant star with a radius of 4.94 times that of the Sun (4.94\,\( R_{\odot} \)). Considering its evolutionary stage, it is likely that the surface composition of the star, even if initially enhanced by accretion, closely resembles its original metallicity at birth. While detecting these enhanced surface compositions from evolved companion stars may be challenging, we predict that such enhancements will be relatively common in low-metallicity, sun-like stars that orbit massive black holes. However, these enhancements are expected to be minimal for lower mass stars that exhibit full convection. As we discover more of these binary systems, it is essential to understand how the observed metallicity can significantly differ from the metallicity at birth.

\subsection{Gaia BH3-like binaries
in Globular Clusters}
Numerous stellar-mass black hole candidates in binary systems have been successfully identified within Galactic globular clusters \citep{1993Natur.364..421K, 2012Natur.490...71S, 2017MNRAS.467.2199B, 2018MNRAS.475L..15G,2022NatAs...6.1085S}.
These discoveries are particularly exciting given the significant advancements in the field of gravitational waves in recent years, which suggest that dynamic interactions in dense star clusters could lead to black hole binary mergers \citep{2014ApJ...784...71S,2016PhRvD..93h4029R,2018ApJ...864...13W,  2020ApJ...903...45K,2024ApJ...975...77Y}. 

Using a typical initial mass function, a globular cluster containing a million initial stars could produce thousands of black holes, many of which are expected to be in binary systems \citep[e.g.,][]{2023ApJ...946L...2V}. If a black hole–hosting binary receives a natal kick exceeding the stellar cluster’s escape velocity, it will be ejected, and its subsequent mass accretion history will follow the scenario described in Section~\ref{sec:dis_ms_bh3}. On the other hand, if they receive a natal kick smaller than the cluster's escape velocity, they will be retained. In this case, the mass accretion history may be further enhanced by gas retention and prolonged enrichment from evolved populations of cluster stars \citep{2007A&A...470..179P, 2008MNRAS.385.2034V, 2009MNRAS.397..488P, 2011ApJ...741...72C, 2018MNRAS.478.2794N}.

In this context, Gaia BH3, linked to the ED-2 retrograde halo stellar stream, is believed to have originated from a disrupted star cluster \citep{2024A&A...687L...3B,2024A&A...688L...2M}. As a result, it may have undergone additional mass accretion from the retained ejecta of asymptotic giant branch stars within the cluster \citep{2018MNRAS.478.2794N}. This process could have altered the companion's abundance patterns for elements such as helium (He), magnesium (Mg), carbon (C), nitrogen (N), oxygen (O), sodium (Na), and aluminum (Al) \citep[e.g.,][]{2008A&A...479..805V}. The accretion history of stars with black hole companions in stellar clusters is certainly more complex than described in Section~\ref{sec:dis_ms_bh3}, and its associated chemical signatures could provide further insights into their origins. Future research will assess the long-term impact of accretion on binary systems in globular clusters.

\section{Summary and Conclusions}\label{sec: conclusions}
We present here an analytical framework designed to quantify the enhanced accretion of metals onto low-mass companions of black holes. To validate our analytical model, we conduct a series of hydrodynamical simulations, which help us develop a useful fitting formula. These simulations reveal how the accretion rate of a low-mass star varies based on the mass of the black hole companion and the binary separation. 

In particular, we examine the black hole hosting binary system known as \textit{Gaia} BH3, focusing specifically on the accretion of iron throughout the evolutionary history of the low-mass companion star. This case is especially intriguing due to the star's low metal abundance. Our findings indicate that such systems have the potential to experience significant metal enhancement through accretion over their lifetimes. However, the signatures of ISM contamination become significantly diluted during the post-main-sequence evolution of these low-mass stars. Some of the key findings of this study are as follows:

\begin{itemize}
    \item We use hydrodynamical simulations to study the mass accretion rate around the sink particles and the density structures that develop in the vicinity of the binary system  (Figure~\ref{fig: dens-redist}). This allows us to estimate the average local density surrounding the less massive stellar companion, which is used to estimate its mass accretion rate. By following this procedure, we derive an expression for how the mass accretion rate onto the less massive companion varies with the mass ratio and the separation of the binary components. Specifically we find that $\dot{M_{*,\rm{bin}}} \propto  (M_{\rm{BH}}/M_*) (R_a/a)$.
    
    \item We show, using the stellar particle trajectories from the {\it Eris} simulation, that the companion’s metal accretion history from formation to today, when enhanced by a massive black hole companion, can play a prominent role in increasing their birth surface abundances (Figure~\ref{fig:redshift}). 
    
    \item A major consequence of ISM pollution is the possible difference between the true birth metallicity of the system and the metallicity determined from spectroscopic observations.
    
    \item We conclude that in binary systems with lighter black hole companions that are widely separated, the resulting surface abundances are low and closely resemble the star’s original metallicity at its formation. In contrast, for binaries with high-mass black hole companions that are in close proximity, the surface abundance resulting from metal accretion can be significant (Figure~\ref{fig: a_m_z_plot}). However, this effect will likely only be measurable in stars that are in the main sequence.  

    \item When a stellar companion evolves off the main sequence, as seen with the \textit{Gaia} BH3 companion, the initial dredge-up process can significantly decrease any excess surface abundance created from the accreted metals (Figure~\ref{fig: z_plot}). Considering its evolutionary stage, the surface composition of the \textit{Gaia} BH3 companion should thus closely resembles its original birth metallicity. 
\end{itemize}

\begin{acknowledgements}
We thank the referee for their insightful comments, which have significantly improved the clarity of this paper.
Our views on the topics discussed herein have been clarified through insightful exchanges with K.~El-Badry, R.~Foley, B.~Gaensler, and E.~Kirby. We acknowledge use of the lux supercomputer at UC Santa Cruz, funded by NSF MRI grant AST 1828315.  We acknowledge support by the Heising-Simons Foundation and the NSF (AST-2307710, AST-2206243, AST-1911206, and AST-1852393). The 3D hydrodynamics software used in this work was developed in part by the DOE NNSA- and DOE Office of Science-supported FLASH Center for Computational Science at the University of Chicago and the University of Rochester. SS acknowledges support from the European High Performance Computing Joint Undertaking (EuroHPC JU) and the Research Council of Norway through the funding of the SPACE Center of Excellence (grant agreement N0101093441).
MSF gratefully acknowledges support provided by the Office of the Vice Chancellor for Research and Graduate Education at the University of Wisconsin--Madison with funding from the Wisconsin Alumni Research Foundation.

\end{acknowledgements}

\software{FLASH 4.3 \citep{Fryxell2000}, MESA \citep{Buchler1976,Fuller1985,Iglesias1993,Oda1994,Saumon1995,Iglesias1996,Itoh1996,Langanke2000,Timmes2000,Rogers2002,Irwin2004,Ferguson2005,Cassisi2007,Chugunov2007,Cyburt2010,Potekhin2010,Paxton2011,Paxton2013,Paxton2015,Paxton2018,Paxton2019,Jermyn2023}, MESA SDK \citep{Townsend2021}, yt \citep{Turk2011}}

\bibliography{binaries.bib,bib.bib}
\bibliographystyle{aasjournalv7}

\begin{appendix}

\section{Stellar Velocity}\label{app:velocity}
\begin{figure}[h]
\centering
\includegraphics[width=0.48\textwidth]{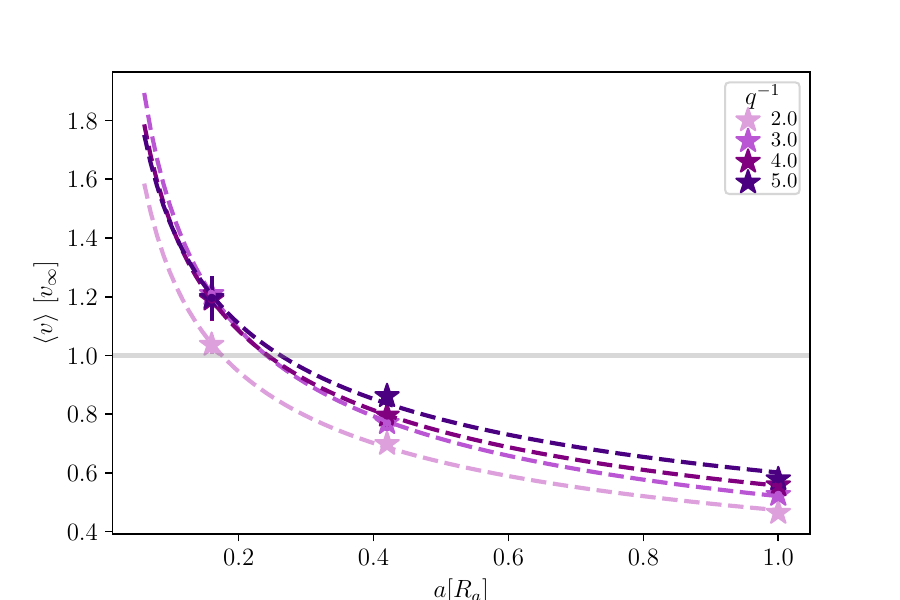}
\caption{Orbit averaged velocity of the star (in units of $v_{\infty}$) as a function of initial separation $a$. The gray horizontal line represents where the orbital velocity is equal to $v_{\infty}$. The colors represent the different mass ratios, where the darker colors are more extreme mass ratios, where $q^{-1} = M_{\rm{BH}}/M_*$. 
}
\label{fig: v_trend}
\end{figure}

When calculating the mass accretion rate of a star in a binary system, two velocities are considered: the orbital velocity of the star around the black hole, denoted as $v_{*, \rm{orb}}$, and the velocity of the center of mass of the system relative to the background gas, represented as $v_{\infty}$. In a system with no external forces, $v_{\rm{orb}}$ is defined to be, $v_{*,\rm{orb}} = \sqrt{GM_{\rm{BH}}/a}$. The center of mass velocity $v_{\infty}$ will be equal to $v_{*,\rm{orb}}$ at a separation $a/R_a = 0.5$. Therefore, we would expect the average orbital velocity, $\langle v \rangle$, to equal  $v_{\infty}$ at $a = 0.5 R_a$. However, numerical results indicate that this is not the case. In Figure~\ref{fig: v_trend}, we plot the numerical solution for the average velocity of the star. We observe that at \( a = 0.5 R_a \), the velocity of the star falls well below the analytical value of \( 1.0 v_{\infty} \). This observation leads us to conclude that, for most separations, the dominant term in the velocity is \( v_{\infty} \).  

For very close companion stars, the orbital velocity becomes more significant, making it more challenging for the companion star to accrete material. At high orbital velocities, the accretion radius approaches the geometric radius of the star, which limits the effectiveness of the Bondi accretion model.

\section{Density redistribution based on adiabatic index}\label{app:gamma}
In dense environments, shortly after a binary system forms, cooling is anticipated to play a significant role \citep[e.g.,][]{2019ApJ...876..142K,2021ApJ...917...36K}. In these situations, the material becomes more compressible, allowing for more effective accretion. Consequently, the gas can be regarded as nearly isothermal rather than adiabatic. To approximate the effects of the radiative cooling on the gas, we vary the adiabatic constant. For the nearly isothermal case, we use $\gamma = 1.1$. We also explore  $\gamma = 4/3$ for completeness.     

\begin{figure}[h]
\centering
\includegraphics[width=0.44\textwidth]{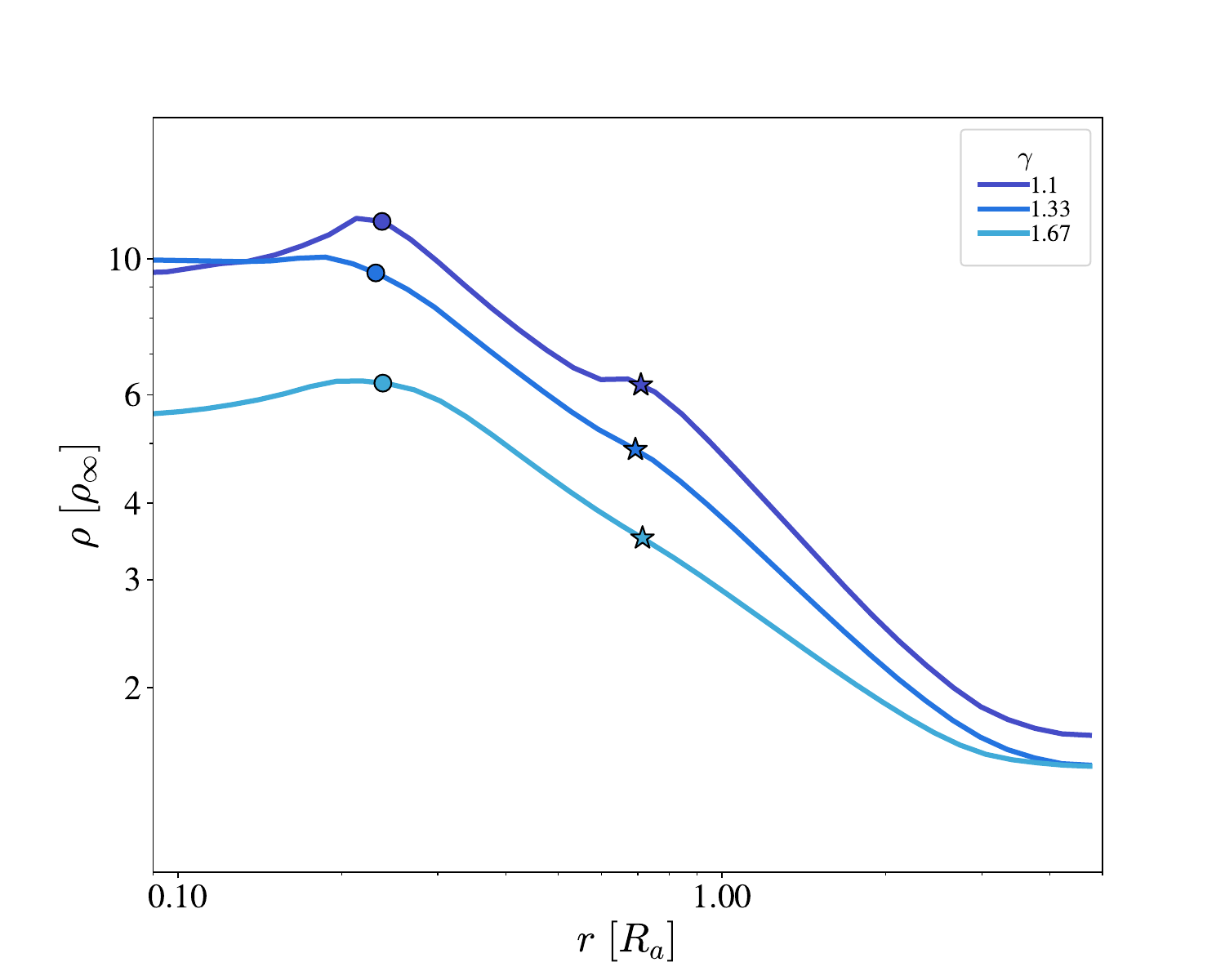}
\caption{Density profiles from the center of mass of the binary obtained  using  varying  gas adiabatic indices at $a =0.42R_a$ and $q=1/3$. From top to bottom, $\gamma=1.1, 4/3,5/3$. The circles mark the locations of the black holes and the star shaped symbols mark the locations of the companion stars relative to the center of mass of the binary. Overall, stars experience an increase in mass accretion rate as gas becomes progressively more compressible. 
}
\label{fig: gamma_change}
\end{figure}

In Figure~\ref{fig: gamma_change}, we present the density profile for three different adiabatic indices with parameters set to $a = 0.42R_a$ and $q=1/3$. The overall trend indicates an increase in density throughout the entire profile as the adiabatic index $\gamma$ decreases. This behavior is expected; in the quasi-isothermal case where $\gamma=1.1$, the gas cools effectively, allowing the material to compress around the black hole, resulting in a higher density cusp. In contrast, with higher values of $\gamma$ , the gas heats up, which inhibits the inflow of material towards the black hole and leads to a lower density throughout the shock region.

The increase in density around the black hole directly contributes to the accretion of metal onto the companion star. Figure~\ref{fig: gamma_change} illustrates a twofold increase in the background density experienced by the star. Since the mass accretion rate $\dot{M}_\ast$  is proportional to the relative density as shown in Equation \ref{eq: mdot}, this rise in background metallicity has the potential to double the star's accretion rate. As a result, this allows for an increase in the total amount of metals being accumulated.

\section{Black hole feedback}\label{app:feedback}
{One of the main uncertainties in the accretion flow surrounding the companion is the role of feedback from the BH companion. Feedback can be deposited into the gas either radiatively or mechanically, through winds, and can affect the mass accretion rate onto the stellar companion. At the low mass accretion rates relevant to our system, accretion is expected to occur radiatively inefficiently. In such cases, known as the adiabatic inflow–outflow solution \citep[ADIOS;][]{1999MNRAS.303L...1B,2004MNRAS.349...68B}, accretion tends to be strongly inhibited, with much less gas reaching the black hole than is supplied at large distances. This is because the outward transport of angular momentum also involves the transport of energy, which can cause gas far from the black hole to become unbound. }

{For perfectly adiabatic (non-radiating) systems, the mass inflow rate depends on radius as $\dot{M}_w \propto R$, while the wind kinetic luminosity scales as $L_w \propto R^{1/2}$. This implies that most of the wind’s kinetic energy is generated at large radii \citep{2012MNRAS.420.2912B}:
\begin{equation}
\begin{split}
    L_w &\approx \dot{M}_w v_w^2 \\
    & \approx \eta \dot{M}_{\rm{B}}v_w^2,
\end{split}
\end{equation}
where $\dot{M}_w$ is the mass loss from the black hole due to winds and $v_w$ is the velocity of the ejected wind. Here 
$\eta = R_{\rm{disk}}/R_a$ and $R_{\rm{disk}}$ is the characteristic radius at which the infalling gas, conserving its angular momentum, begins to form a stable, approximately Keplerian disk.}

{To determine the value of $R_{\rm{disk}}$, we estimate the circularization radius based on the distance between the black hole and the first Lagrangian point of the binary system, $R_{L1}$. This distance can be approximated as}
\begin{equation}
     R_{L1} \approx a \left( 0.5 + 0.277\log q\right).
\end{equation}

The disk radius is then defined as, 
\begin{equation}
    R_{\rm{disk}} \approx a(1+q)\left( \frac{R_{L1}}{a} \right)^4.
\end{equation}
Using this definition, we can find the value of $\eta$, 
\begin{equation}
    \eta = {R_{\rm disk} \over R_a}\approx \frac{a}{R_a}(1+q) \left( 0.5 + 0.277\log q\right)^4. 
\end{equation}
{For the \textit{Gaia} BH3 system, $q=1/40$ and $a/R_a$ varies with time. The value of $\eta$ will be greatest at larger separations. We therefore take $a/R_a = 1$ to represent the value with the greatest disk extension that still falls within the regime of binary accretion.  Using this value and the mass ratio of approximately 40 for Gaia BH3, we find $\eta = 3.5 \times 10^{-4}$.}

{If an accreting BH produces a wind, it can potentially carve out a cavity in the surrounding gas, reducing or even suppressing the gas supply available for accretion. We calculate the shock radius of this cavity, $R_s$, as the location where the ram pressure of the wind balances the ambient gas pressure,  which can be expressed as
\begin{equation}
    R_s = \sqrt{\frac{\dot{M}_w v_w}{4 \pi \rho_{\infty} v_{\infty}^2} }.
\end{equation}
To compare this to the relevant physical scale, we can express the shock radius in terms of the Bondi radius,
\begin{equation}
    R_a = \sqrt{\frac{\dot{M}_{\rm{B}}}{\pi \rho_{\infty} v_\infty}}.
\end{equation}
This ratio can then be written as
\begin{equation}
\begin{split}
    \frac{R_s}{R_a}& = \sqrt{\frac{\dot{M}_w v_w}{4\dot{M}_{\rm{B}} v_\infty}} \\
    &= \frac{1}{2} \sqrt{\frac{\dot{M}_w}{\dot{M}_{\rm{B}}}} \sqrt{\frac{v_w}{v_\infty}},
\end{split}  
\end{equation}
where $\dot{M}_w =\eta \dot{M}_{\rm B}$ and $\frac{v_w}{v_\infty}\approx \eta^{-1/2}$, so that 
\begin{equation}
     \frac{R_s}{R_a} \approx \frac{1}{2}\eta^{1/4},
\end{equation}
Here, $\eta \ll 1$. This suggests that for radiatively inefficient accretion flows, the luminosity produced by the black hole is unlikely to carve out a large cavity or significantly impede the gas supply onto the companion. Nevertheless, the flow may still be reshaped, potentially affecting the scaling laws derived in this paper (Section~\ref{sec: results}), which were obtained under the assumption of no feedback. }

\end{appendix}

\end{document}